# Budget Optimization in Search-Based Advertising Auctions

Jon Feldman [*]    S. Muthukrishnan [†]    Martin Pál [‡]    Cliff Stein [§]

February 1, 2008

## Abstract

Internet search companies sell advertisement slots based on users' search queries via an auction. While there has been a lot of attention on the auction process and its game-theoretic aspects, our focus is on the advertisers. In particular, the advertisers have to solve a complex optimization problem of how to place bids on the keywords of their interest so that they can maximize their return (the number of user clicks on their ads) for a given budget. We model the entire process and study this budget optimization problem. While most variants are NP hard, we show, perhaps surprisingly, that simply randomizing between two uniform strategies that bid equally on all the keywords works well. More precisely, this strategy gets at least $1 - 1/e$ fraction of the maximum clicks possible. Such uniform strategies are likely to be practical. We also present inapproximability results, and optimal algorithms for variants of the budget optimization problem.

## 1 Introduction

Online search is now ubiquitous and Internet search companies such as Google, Yahoo! and MSN let companies and individuals advertise based on search queries posed by users. Conventional media outlets, such as TV stations or newspapers, price their ad slots individually, and the advertisers buy the ones they can afford. In contrast, Internet search companies find it difficult to set a price explicitly for the advertisements they place in response to user queries. This difficulty arises because supply (and demand) varies widely and unpredictably across the user queries, which range from *specialty vacations* to *mundane home supplies*; furthermore they have to price slots for billions of such queries in real time. Instead, they rely on the market to determine suitable prices by using auctions amongst the advertisers. It is a challenging problem to set up the auction in order to effect a stable market in which all the parties (the advertisers, users as well as the Internet search company) are adequately satisfied. Recently there has been systematic study of the issues involved in the game theory of the auctions [4, 1, 2], revenue maximization [9], etc.

The perspective in this paper is not of the Internet search company that displays the advertisements, but rather of the advertisers. The challenge from an advertiser's point of view is to understand and interact with the auction mechanism. The advertiser determines a set of keywords of their interest and then must create ads, set the bids for each keyword, and provide a total (often daily) budget. When a user poses a search query, the Internet search company determines the advertisers whose keywords match the query and who still have budget left over, runs an auction amongst them, and presents the set of ads corresponding to the advertisers who "win" the auction. The advertiser whose ad appears pays the Internet search company if the user clicks on the ad.

The focus in this paper is on how the advertisers bid. For the particular choice of keywords of their interest[1], an advertiser wants to optimize the overall effect of the advertising campaign. While the effect of an ad campaign in any medium is a complicated phenomenon to quantify, one commonly accepted (and easily quantified) notion in search-based advertising on the Internet is to maximize the number of clicks. The Internet search companies are supportive towards advertisers and provide statistics about the history of click volumes and prediction about the future performance of various keywords. Still, this is a complex problem for the following reasons (among others):

- Individual keywords have significantly different characteristics from each other; e.g., while "fishing" is a broad keyword that matches many user queries and has many competing advertisers, "humane fishing bait" is a niche keyword that matches only a few queries, but might have less competition.

- There are complex *interactions* between keywords because a user query may match two or more keywords. This is quite common since the advertiser is trying to cover all the possible keywords in some domain. In effect the advertiser ends up competing with themselves.

---

[*]Google, Inc., New York, NY. jonfeld@google.com
[†]Google, Inc., New York, NY. muthu@google.com
[‡]Google, Inc., New York, NY. mpal@google.com
[§]Department of IEOR, Columbia University. cliff@ieor.columbia.edu. This work was done while visiting Google, Inc., New York, NY.

[1]The choice of keywords is related to the domain-knowledge of the advertiser, user behavior and strategic considerations. Internet search companies provide the advertisers with summaries of the query traffic which is useful for them to optimize their keyword choices interactively. We do not directly address the choice of keywords in this paper, which is addressed elsewhere [11].

As a result, the advertisers face a challenging optimization problem. The focus of this paper is to solve this optimization problem.

**1.1 The Budget Optimization Problem.** We present a short discussion and formulation of the optimization problem faced by advertisers; a more detailed description is in Section 2.

A given advertiser sees the state of the auctions for search-based advertising as follows. There is a set $K$ of keywords of interest; in practice, even small advertisers typically have a large set $K$. There is a set $Q$ of queries posed by the users. For each query $q \in Q$, there are functions giving the $\text{clicks}_q(b)$ and $\text{cost}_q(b)$ that result from bidding a particular amount $b$ in the auction for that query, which we model more formally in the next section. There is a bipartite graph $G$ on the two vertex sets representing $K$ and $Q$ respectively. For any query $q \in Q$, the neighbors of $q$ in $K$ are the keywords that are said to "match" the query $q$.[2] The *budget optimization problem* is as follows. Given graph $G$ together with the functions $\text{clicks}_q(\cdot)$ and $\text{cost}_q(\cdot)$ on the queries, as well as a budget $U$, determine the assignment of bids $b_k$ to each keyword $k \in K$ such that $\sum_q \text{clicks}_q(b_q)$ is maximized subject to $\sum_q \text{cost}_q(b_q) \leq U$, where the "effective bid" $b_q$ on a query is some function of the keyword bids in the neighborhood of $q$.

While we can cast this problem as a traditional optimization problem, there are different challenges in practice depending on the advertiser's access to the query and graph information, and indeed the reliability of this information (e.g., it could be based on unstable historical data). Thus it is important to find solutions to this problem that not only get many clicks, but are also simple, robust and less reliant on the information. In this paper we define the notion of a "uniform" strategy which is essentially a strategy that bids uniformly on all keywords. Since this type of strategy obviates the need to know anything about the particulars of the graph, and effectively aggregates the click and cost functions on the queries, it is quite robust, and thus desirable in practice. What is surprising is that uniform strategy actually performs well, which we will prove.

**1.2 Our Main Results and Technical Overview.** We present positive and negative results for the budget optimization problem. In particular, we show:

- Nearly all formulations of the problem are NP-Hard. In cases slightly more general than the formulation above, where the clicks have weights, the problem is inapproximable better than a factor of $1 - \frac{1}{e}$.

- We give a $(1 - 1/e)$-approximation algorithm for the budget optimization problem. The strategy found by the algorithm is a *two-bid uniform strategy*, which means that it randomizes between bidding some value $b_1$ on all keywords, and bidding some other value $b_2$ on all keywords.[3] We show that this approximation ratio is tight for uniform strategies.

  We also give a $(1/2)$-approximation algorithm that offers a *single-bid uniform strategy*, only using one value $b_1$. (This is tight for single-bid strategies.) These strategies can be computed in time nearly linear in $|Q| + |K|$, the input size.

Uniform strategies may appear to be naive in first consideration because the keywords vary significantly in their click and cost functions, and there may be complex interaction between them when multiple keywords are relevant to a query. After all, the optimum can configure arbitrary bids on each of the keywords. Even for the simple case when the graph is a *matching*, the optimal algorithm involves placing different bids on different keywords as given by a walk on the "convex hulls" of the functions associated with the keywords (Section B). So, it might be surprising that a simple two-bid uniform strategy is 63% or more effective compared to the optimum. In fact, our proof is stronger, showing that this strategy is 63% effective against a strictly more powerful adversary who can bid independently on the *individual queries*, i.e., not be constrained by the interaction provided by the graph $G$.

---

[2]The particulars of the matching rule are determined by the Internet search company; here we treat the function as arbitrary.

[3]This type of strategy can also be interpreted as bidding one value (on all keywords) for part of the day, and a different value for the rest of the day.

As in nearly all the proofs that generate a $1 - 1/e$ approximation ratio, an adversarial analysis and a distribution over strategies play a key role, and we discovered the optimal distributions using a factor-revealing LP as described in Section 4.

Without overemphasizing, it should be clear that advertisers might have a preference for uniform bidding strategies. Our preliminary experimental results using real auction data logs at Google indicate that these simple bidding strategies are significantly better than the theoretical guarantees, but the focus here remains proving theoretical bounds.

The formulation of the problem above and its justification is presented in Section 2 with a detailed modeling of the auction process in Internet search companies. We then define precisely the notion of a uniform strategy (Section 3), and present our approximation algorithm (Section 4). We conclude with a discussion of extensions and open problems. In the appendix we present hardness results for the budget optimization problem and its variants, and give exact algorithms for some special cases.

## 2 Modeling a Keyword Auction

We describe an auction from an advertiser's point of view. An advertiser bids on a *keyword,* which we can think of as a word or set of words. Users of the search engine submit *queries*. If the query "matches" a keyword that has been bid on by an advertiser, then the advertiser is entered into an auction for the available ad slots on the results page. What constitutes a "match" varies depending on the search engine.

Note that an advertiser makes a single bid for a keyword that remains in effect for a period of time, say one day, and the keyword could match many different user queries throughout the day. Each user query might have a different set of advertisers competing for clicks. The advertiser could also bid different amounts on multiple keywords, each matching a (possibly overlapping) set of user queries.

The ultimate goal of an advertiser is to maximize traffic to their website, given a certain advertising budget. We now formalize a model of keyword bidding and define an optimization problem that captures this goal.

**2.1 Landscapes.** We begin by considering the case of a *single* keyword that matches a *single* user query. In this section we define the notion of a "query landscape" that describes the relationship between the advertiser's bid and what will happen on this query as a result of this bid.[8] This definition will be central to the discussion as we continue to more general cases.

**2.1.1 Positions, bids and click-through rate.** The search results page for a query contains $p$ possible positions in which our ad can appear. We denote the highest (most favorable) position by 1 and lowest by $p$.

Associated with each position $i$ is a value $\alpha[i]$ that denotes the *click-through rate* (ctr) of the ad in position $i$. The ctr is a measure of how likely it is that our ad will receive a click if placed in position $i$. The ctr can be measured empirically using past history. We assume throughout this work that that $\alpha[i] \leq \alpha[j]$ if $j < i$, that is, higher positions receive at least as many clicks as lower positions.

In order to place an ad on this page, we must enter the *auction* that is carried out among all advertisers that have submitted a *bid* on a keyword that matches the user's query. We will refer to such an auction as a *query auction*, to emphasize that there is an auction for each query rather than for each keyword. We assume that the auction is a *generalized second price* (GSP) auction [4, 6]: the advertisers are ranked in decreasing order of bid, and each advertiser is assigned a price equal to the amount bid by the advertiser below them in the ranking.[4] Note that in sponsored search auctions, this advertiser pays only if the user actually clicks on the ad. Let $(b[1], \ldots, b[p])$ denote the bids of the top $p$ advertisers in this query auction. For notational convenience, we assume that $b[0] = \infty$ and $b[p] = \alpha[p] = 0$. Since the auction is a generalized second price auction, higher bids win higher positions; i.e. $b[i] \geq b[i+1]$. Suppose that we bid $b$ on some keyword that matches the user's query, then

---

[4]Google, Yahoo! and MSN all use some variant of the GSP auction. Also, other auctions besides GSP have been considered; e.g., the Vickrey Clark Groves (VCG) auction [12, 3, 6]. Each auction mechanism will result in a different sort of optimization problem. In the conclusion we point out that for the VCG auction, the bidding optimization problem becomes quite easy.

our position is defined by the largest b[i] that is at most b, that is,

$$\text{pos}(b) = \arg\max_i (b[i] : b[i] \leq b). \quad (1)$$

Since we only pay if the user clicks (and that happens with probability $\alpha[i]$), our expected *cost* for winning position $i$ would be $\text{cost}[i] = \alpha[i] \cdot b[i]$, where $i = \text{pos}(b)$. We use $\text{cost}_q(b)$ and $\text{clicks}_q(b)$ to denote the expected cost and clicks that result from having a bid $b$ that qualifies for a query auction $q$, and thus

$$\text{cost}_q(b) = \alpha[i] \cdot b[i] \text{ where } i = \text{pos}(b), \quad (2)$$

$$\text{clicks}_q(b) = \alpha[i] \text{ where } i = \text{pos}(b). \quad (3)$$

The following observations about cost and clicks follow immediately from the definitions and equations (1), (2) and (3).

**Observation 1** *For $b \in \mathbb{R}_+$,*

1. *$(\text{cost}_q(b), \text{clicks}_q(b))$ can only take on one of a finite set of values $V_q = \{(\text{cost}[1], \alpha[1]), \ldots, (\text{cost}[p], \alpha[p])\}$.*

2. *Both $\text{cost}_q(b)$ and $\text{clicks}_q(b)$ are non-decreasing functions of $b$. Also, cost-per-click (cpc) $\text{cost}_q(b)/\text{clicks}_q(b)$ is non-decreasing in $b$.*

3. *$\text{cost}_q(b)/\text{clicks}_q(b) \leq b$.*

For bids $(b[1], \ldots, b[p])$ that correspond to the bids of other advertisers, we have: $\text{cost}_q(b[i])/\text{clicks}_q(b[i]) = b[i]$, $i \in [p]$. When the context is clear, we drop the subscript $q$.

**2.1.2 Landscapes.** We can summarize the data contained in the functions $\text{cost}(b)$ and $\text{clicks}(b)$ as a collection of points in a plot of cost vs. clicks, which we refer to as a *landscape*. For example, for a query with four slots, a landscape might look like Table 1.

| bid range | cost per click | cost | clicks |
|---|---|---|---|
| [$2.60,∞) | $2.60 | $1.30 | .5 |
| [$2.00,$2.60) | $2.00 | $0.90 | .45 |
| [$1.60,$2.00) | $1.60 | $0.40 | .25 |
| [$0.50,$1.60) | $0.50 | $0.10 | .2 |
| [$0,$0.50) | $0 | $0 | 0 |

Table 1: A *landscape* for a query

It is convenient to represent this data graphically as in Figure 1. Here we graph clicks as a function of cost.

Observe that in this graph, the cpc ($\text{cost}(b)/\text{clicks}(b)$) of each point is the reciprocal of the slope of the line from the origin to the point. Since $\text{cost}(b)$, $\text{clicks}(b)$ and $\text{cost}(b)/\text{clicks}(b)$ are non-decreasing, the slope of the line from the origin to successive points on the plot decreases. This condition is slightly weaker than concavity.

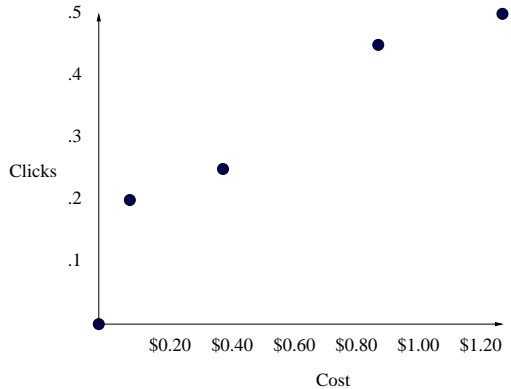

Figure 1: A bid landscape. We graph clicks as a function of cost.

Suppose we would like to solve budget optimization problem for a single keyword landscape.[5] As we increase our bid from zero, our cost increases and our expected number of clicks increases, and so we simply submit the highest bid such that we remain within our budget.

One problem we see right away is that since there are only a finite set of points in this landscape, we may not be able to target arbitrary budgets efficiently. Suppose in the example from Table 1 that we had a budget of $1.00. Bidding between $2.00 and $2.60 uses only $0.90, and so we are underspending. Bidding more than $2.60 is not an option, since we would then incur a cost of $1.30 and overspend our budget.

**2.2 Randomized strategies.** To rectify this problem and better utilize our available budget, we allow *randomized bidding strategies*. Let $\mathcal{B}$ be a distribution on bids $b \in \mathbb{R}_+$.[6] Now we define $\text{cost}(\mathcal{B}) = E_{b \sim \mathcal{B}}[\text{cost}(b)]$ and $\text{clicks}(\mathcal{B}) = E_{b \sim \mathcal{B}}[\text{clicks}(b)]$. Graphically, the possible values of

---

[5]Of course it is a bit unrealistic to imagine that an advertiser would have to worry about a budget if only one user query was being considered; however one could imagine multiple instances of the same query and the problem scales.

[6] We use $\mathbb{R}_+$ to denote the nonnegative reals.

$(\text{cost}(\mathcal{B}), \text{clicks}(\mathcal{B}))$ lie in the convex hull of the landscape points. See Figure 2 for the set of allowable bids for the example in Table 1.

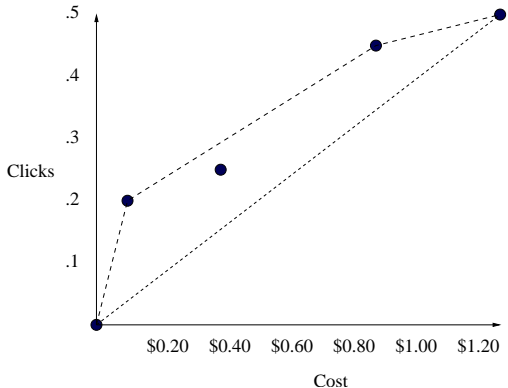

Figure 2: The convex hull of allowable randomized bids.

To find a bid distribution $\mathcal{B}$ that maximizes clicks subject to a budget, we simply draw a vertical line on the plot where the cost is equal to the budget, and find the highest point on this line in the convex hull. This point will always be the convex combination of at most *two* original landscape points which themselves lie *on* the convex hull. Thus, given the point on the convex hull, it is easy to compute a distribution on two bids which led to this point. Summarizing,

**Lemma 1** *If an advertiser is bidding on one keyword, subject to a budget $U$, then the optimal strategy is to pick a convex combination of (at most) two bids which are at the endpoints of the line on the convex hull at the highest point for cost $U$.*

There is one subtlety in this formulation. Given any bidding strategy, randomized or otherwise, the resulting cost is itself a random variable representing the expected cost. Thus if our budget constraint is a hard budget, we have to deal with the difficulties that arise if our strategy would be over budget. Therefore, we think of our budget constraint as *soft*, that is, we only require that our expected cost be less than the budget. In practice, the budget is often an average daily budget, and thus we don't worry if we exceed it one day, as long as we are meeting the budget in expectation. Further, either the advertiser or the search engine (possibly both), monitor the cost incurred over the day; hence, the advertiser's bid can be changed to zero for part of the day, so that the budget is not overspent.[7] Thus in the remainder of this paper, we will formulate a budget constraint that only needs to be respected in expectation.

**2.3 Keyword Interaction.** In reality, search advertisers can bid on a large set of keywords, each of them qualifying for a different (possibly overlapping) set of queries.

One complicating factor is that most search engines do not allow an advertiser to appear twice in the same search results page.[8] Thus, if an advertiser has a bid on two different keywords that match the same query, this conflict must be resolved somehow. For example, if an advertiser has a bid out on the keywords "shoes" and "high-heel," then if a user issues the query "high-heel shoes," it will match on two different keywords. The search engine specifies, in advance, a rule for resolution based on the query the keyword and the bid. The most natural rule is to take the keyword with the highest bid, which we adopt here, but our results apply to other natural resolution rules.

We model this problem using an undirected bipartite graph $G = (K \cup Q, E)$ where $K$ is a set of keywords and $Q$ is a set of queries. Each $q \in Q$ has an associated landscape, as defined by $\text{cost}_q(b)$ and $\text{clicks}_q(b)$. An edge $(k, q) \in E$ means that keyword $k$ matches query $q$.

The advertiser can control their individual *keyword bid vector* $\mathbf{a} \in \mathbb{R}_+^{|K|}$ specifying a bid $\mathbf{a}_k$ for each keyword $k \in K$. (For now, we do not consider randomized bids, but we will introduce that shortly.) Given a particular bid vector $\mathbf{a}$ on the keywords, we use the resolution rule of taking the maximum to define the "effective bid" on query $q$ as

$$\mathrm{b}_q(\mathbf{a}) = \max_{k:(k,q)\in E} \mathbf{a}_k.$$

By submitting a bid vector $\mathbf{a}$, the advertiser receives some number of clicks and pays some cost on each keyword. We use the term *spend* to denote the total cost, i.e. the amount spent. Similarly, we use the term *traffic* to denote the total number of clicks. These are defined by the following formulae.

$$\begin{aligned} \text{spend}(\mathbf{a}) &= \sum_{q\in Q} \text{cost}_q(\mathrm{b}_q(\mathbf{a})) \\ \text{traffic}(\mathbf{a}) &= \sum_{q\in Q} \text{clicks}_q(\mathrm{b}_q(\mathbf{a})) \end{aligned}$$

---

[7]See https://adwords.google.com/support/bin/answer.py?answer=22183, for example.

[8]See https://adwords.google.com/support/bin/answer.py?answer=14179, for example.

We also allow randomized strategies, where an advertiser gives a distribution $\mathcal{A}$ over bid vectors $\mathbf{a} \in \mathbb{R}_+^{|K|}$. The resulting spend and traffic are given by

$$\text{spend}(\mathcal{A}) = E_{\mathbf{a} \sim \mathcal{A}}[\text{spend}(\mathbf{a})],$$

$$\text{traffic}(\mathcal{A}) = E_{\mathbf{a} \sim \mathcal{A}}[\text{traffic}(\mathbf{a})].$$

We can now state the BUDGET OPTIMIZATION problem in its full generality:

BUDGET OPTIMIZATION
**Input:** a budget $U$, a keyword-query graph $G = (K \cup Q, E)$, and landscapes $(\text{cost}_q(\cdot), \text{clicks}_q(\cdot))$ for each $q \in Q$.
**Find:** a distribution $\mathcal{A}$ over bid vectors $\mathbf{a} \in \mathbb{R}_+^{|K|}$ such that $\text{spend}(\mathcal{A}) \leq U$ and $\text{traffic}(\mathcal{A})$ is maximized.

We conclude this section with a small example to illustrate the complications in this problem. Suppose you have two keywords $K = \{u, v\}$ and two queries $Q = \{x, y\}$ and edges $E = \{(u, x), (u, y), (v, y)\}$. Suppose query $x$ has one position with ctr $\alpha^x[1] = 1.0$, and there is one bid $b_1^x = \$1$. Query $y$ has two positions with ctrs $\alpha^y[1] = \alpha^y[2] = 1.0$, and bids $b_1^y = \$\epsilon$ and $b_2^y = \$1$ To get any clicks from $x$, an advertiser must bid at least \$1 on $u$. However, because of the structure of the graph, if the advertiser sets $b_u$ to \$1, then his effective bid is \$1 on both $x$ *and* $y$. Thus he must tradeoff between getting the clicks from $x$ and getting the bargain of a click for $\$\epsilon$ that would be possible otherwise.

## 3 Uniform Bidding Strategies

As we will show in Section C, solving the BUDGET OPTIMIZATION problem in its full generality is difficult. In addition, it may be difficult to reason strategies that involve arbitrary distributions over arbitrary bid vectors. Advertisers generally prefer strategies that are easy to understand, evaluate and use within their larger goals. With this motivation, we look at restricted classes of strategies that we can easily compute, explain and analyze.

We define a *uniform bidding strategy* to be a distribution $\mathcal{A}$ over bid vectors $\mathbf{a} \in \mathbb{R}_+^{|K|}$ where each bid vector in the distribution is of the form $(b, b, \ldots, b)$ for some real-valued bid $b$. In other words, each vector in the distribution bids the same value on every keyword.

Uniform strategies have several advantages. First, they do not depend on the edges of the interaction graph, since all effective bids on queries are the same. Thus, they are effective in the face of limited or noisy information about the keyword interaction graph. Second, uniform strategies are also independent of the priority rule being used. Third, any algorithm that gives an approximation guarantee will then be valid for *any* interaction graph over those keywords and queries.

We now show that we can compute the best *uniform strategy* efficiently.

Suppose we have a set of queries $Q$, where the landscape $V_q$ for each query $q$ is defined by the set of points $V_q = \{(\text{cost}_q[1], \alpha_q[1]), \ldots, (\text{cost}_q[p], \alpha_q[p])\}$. We define the set of *interesting bids* $I_q = \{\text{cost}_q[1]/\alpha_q[1], \ldots, \text{cost}_q[p]/\alpha_q[p]\}$, let $\mathcal{I} = \cup_{q \in Q} I_q$, and let $N = |\mathcal{I}|$. We can index the points in $\mathcal{I}$ as $b_1, \ldots, b_N$ in increasing order. The $i$th point in our *aggregate landscape* $\mathcal{V}$ is found by summing, over the queries, the cost and clicks associated with bid $b_i$, that is, $\mathcal{V} = \cup_{i=1}^N (\sum_{q \in Q} \text{cost}_q(b_i), \sum_{q \in Q} \text{clicks}_q(b_i))$.

For any possible bid $b$, if we use the aggregate landscape just as we would a regular landscape, we exactly represent the cost and clicks associated with making that bid simultaneously on all queries associated with the aggregate landscape. Therefore, all the definitions and results of Section 2 about landscapes can be extended to aggregate landscapes, and we can apply Lemma 1 to compute the best uniform strategy (using the convex hull of the points in this aggregate landscape). The running time is dominated by the time to compute the convex hull, which is $O(N \log N)$[10].

Note also that the strategy we get is the convex combination of two points on the aggregate landscape. Define a *two-bid strategy* to be a uniform strategy which puts non-zero weight on at most two bid vectors. We have shown

**Lemma 2** *Given an instance of the* BUDGET OPTIMIZATION *problem in which there are a total of $N$ points in all the landscapes, we can find the best uniform strategy in $O(N \log N)$ time. Furthermore, this strategy will always be a two-bid strategy.*

Putting these ideas together, we get the following $O(N \log N)$-time algorithm for the BUDGET OPTIMIZATION problem, where $N$ is the total number of landscape points. (We later show that this is a $(1-\frac{1}{e})$-approximation algorithm.)

1. Aggregate all the points from the individual query landscapes into a single aggregate landscape.
2. Find the convex hull of the points in the aggregate landscape.
3. Compute the point on the convex hull for the given budget, which is the convex combination of two points $\alpha$ and $\beta$.
4. Output the strategy which is the appropriate convex combination of the uniform bid vectors corresponding to $\alpha$ and $\beta$.

We will also consider a special case of two-bid strategies. A *single-bid* strategy is a uniform strategy which puts non-zero weight on at most one *non-zero* vector, i.e. advertiser randomizes between bidding a certain amount $b^*$ on all keywords, and not bidding at all. A single-bid strategy is even easier to implement in practice than a two-bid strategy. For example, the search engines often allow advertisers to set a maximum daily budget. In this case, the advertiser would simply bid $b^*$ until her budget runs out, and the ad serving system would remove her from all subsequent auctions until the end of the day. One could also use an "ad scheduling" tool offered by some search companies[9] to implement this strategy. The best single-bid strategy can also be computed easily from the aggregate landscape. The optimal strategy for a budget $U$ will either be the point $x$ s.t. $\text{cost}(x)$ is as large as possible without exceeding $U$, or a convex combination of zero and the point $y$, where $\text{cost}(y)$ is as small as possible while larger than $U$.

## 4 Approximation Algorithms

In the previous section we proposed using uniform strategies and gave an efficient algorithm to compute the best such strategy. In section we prove that there is always a good uniform strategy:

**Theorem 3** *There always exists a uniform bidding strategy that is $(1-\frac{1}{e})$-optimal. Furthermore, for any $\epsilon > 0$, there exists an instance for which all uniform strategies are at most $(1-\frac{1}{e}+\epsilon)$-optimal.*

We introduce the notion of a *click-price curve*, which is central to our analysis. This definition makes it simple to show that there is always a *single-bid* strategy that is a $\frac{1}{2}$-approximation (and this is tight); we then build on this to prove Theorem 3.

---
[9]See https://adwords.google.com/support/bin/answer.py?answer=33227 for example.

| Query | clicks | cost | cpc |
|-------|--------|------|------|
| A | 2 | $1 | $0.50 |
| B | 5 | $0.50 | $0.10 |
| C | 3 | $2 | $0.67 |
| D | 4 | $1 | $0.25 |

Table 2: A set of four queries and their optimal clicks and cost

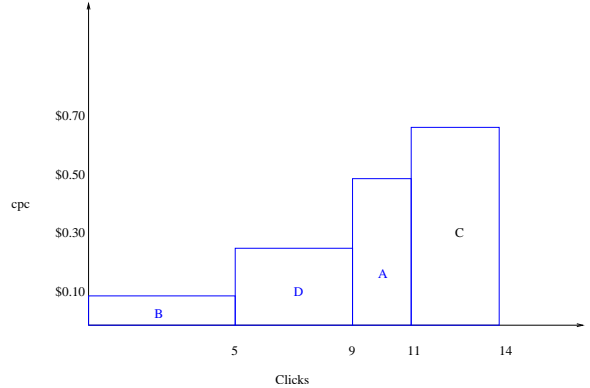

Figure 3: A click-price curve for the queries in Table 2.

**4.1 Click-price curves.** Consider a set of queries $Q$, and for each query $q \in Q$, let $(\text{clicks}_q(\cdot), \text{cost}_q(\cdot))$ be the corresponding bid landscape. Consider an adversarial bidder $\Omega$ with the power to bid independently on each *query*. Note that this bidder is a more powerful than an optimal bidder, which has to bid on the keywords. Suppose this strategy bids $b_q^*$ for each query $q$. Thus, $\Omega$ achieves traffic $C_\Omega = \sum_i \text{clicks}(b_i^*)$, and incurs total spend $U_\Omega = \sum_i \text{cost}(b_i^*)$.

Note that without loss of generality we can assume that $\Omega$ bids so that for each query $q$, the cost per click is equal to $b_q^*$, i.e. $\text{cost}_q(b_q^*)/\text{clicks}_q(b_q^*) = b_q^*$. We may assume this because for some query $q$, if $\text{cost}_q(b_q^*)/\text{clicks}_q(b_q^*) < b_q^*$, we can always lower $b_q^*$ and keep the same cost and clicks.

To aid our discussion, we introduce the notion of a *click-price curve*, an example of which is shown in Figure 3. Formally the curve is a non-decreasing function $h : [0, C_\Omega] \mapsto \mathbb{R}_+$. defined as

$$h(r) = \min\{c \mid \sum_{q: b_q^* \leq c} \text{clicks}_q \geq r\}$$

Another way to construct this curve is to sort the queries in increasing order by $b_q^* = \text{cost}_q(b_q^*)/\text{clicks}_q(b_q^*)$, then make a step function where the $q$th step has height $b_q^*$ and width $\text{clicks}_q(b_q^*)$ (see Figure 3). Note that the area of each step is $\text{cost}_q(b_q^*)$. The following claim follows immediately:

**Claim 1** $U_\Omega = \int_0^{C_\Omega} h(r) dr.$

Suppose we wanted to buy some fraction $r'/C_\Omega$ of the traffic that $\Omega$ is getting. The click-price curve says that if we bid $h(r')$ on every keyword (and therefore every query), we get at least $r'$ traffic, since this bid would ensure that for all $q$ such that $b_q^* \leq h(r')$ we win as many clicks as $\Omega$. Note that by bidding $h(r')$ on every keyword, we may actually get even more than $r'$ traffic, since for queries $q$ where $b_q^*$ is much less than $h(r')$ we may win more clicks than $\Omega$. However, all of these extra clicks still cost at most $h(r')$ per click.

Thus we see that for any $r' \in [0, C_\Omega]$, if we bid $h(r')$ on every keyword, we receive at least $r'$ traffic at a total spend of at most $h(r')$ per click. Note that by randomizing between bidding zero and bidding $h(r')$, we can receive *exactly* $r'$ traffic at a total spend of at most $r' \cdot h(r')$. We summarize this discussion in the following lemma:

**Lemma 4** *For any $r \in [0, C_\Omega]$, there exists a single-bid strategy that randomizes between bidding $h(r)$ and bidding zero, and this strategy receives exactly $r$ traffic with total spend at most $r \cdot h(r)$.*

Finally, we show that the strategies described in Lemma 4 are essentially the only ones available for the distribution in a uniform strategy, at least for a certain class of instances (the proof can be found in the appendix):

**Lemma 5** *For any $C, U > 0$ and non-decreasing function $f : [0, C] \to \mathbb{R}_+$ such that $\int_0^C f(r) dr = U$, and any small $\epsilon' > 0$, there exists an instance of* BUDGET OPTIMIZATION *with budget $U + \epsilon'$, where the optimal solution achieves $C$ clicks at cost $U + \epsilon'$, and all uniform bidding strategies are convex combinations of single-bid strategies that achieve exactly $r$ clicks at cost exactly $rf(r)$ by bidding $f(r)$ on all keywords.*

**4.2 A $\frac{1}{2}$-approximation algorithm.** Using Lemma 4 we can now show that there is a uniform single-bid strategy that is $\frac{1}{2}$-optimal. In addition to being an interesting result in its own right, it also serves as a warm-up for our main result.

**Theorem 6** *There always exists a uniform single-bid strategy that is $\frac{1}{2}$-optimal. Furthermore, for any $\epsilon > 0$ there exists an instance for which all single-bid strategies are at most $(\frac{1}{2} + \epsilon)$-optimal.*

*Proof:* Applying Lemma 4 with $r = C_\Omega/2$, we see that there is a strategy that achieves traffic $C_\Omega/2$ with spend $C_\Omega/2 \cdot h(C_\Omega/2)$. Now, using the fact that $h$ is a non-decreasing function combined with Claim 1, we have

$$(C_\Omega/2)h(C_\Omega/2) \leq \int_{C_\Omega/2}^{C_\Omega} h(r) dr \leq \int_0^{C_\Omega} h(r) dr = U_\Omega, \quad (4)$$

which shows that we spend at most $U_\Omega$. We conclude that there is a $\frac{1}{2}$-optimal single-bid strategy randomizing between bidding $C_\Omega/2$ and zero.

The second part of the Theorem appears in the Appendix. ∎

**4.3 A $(1 - \frac{1}{e})$-approximation algorithm.** We now prove Theorem 3. The key to the proof is to show that there is a distribution over single-bid strategies from Lemma 4 that obtains at least $(1 - \frac{1}{e})C_\Omega$ clicks. In order to figure out the best distribution, we wrote a linear program that models the behavior of a player who is trying to maximize clicks and an adversary who is trying to create an input that is hard for the player. Then using linear programming duality, we were able to derive both an optimal strategy and a tight instance. After solving the LP numerically, we were also able to see that there is a uniform strategy for the player that always obtains $(1 - \frac{1}{e})C_\Omega$ clicks; and then from the solution were easily able to "guess" the optimal distribution. This methodology is similar to that used in work on factor-revealing LPs [7, 9].

**4.3.1 An LP for finding the worst-case click-price curve.** Consider the adversary's problem of finding a click-price curve for which no uniform bidding strategy can achieve $\alpha C_\Omega$ clicks. Recall that by Lemma 1 we can assume that a uniform strategy randomizes between two bids $u$ and $v$. We also assume that the uniform strategy uses a convex combination of strategies from Lemma 4, which we can assume by Lemma 5. Thus, to achieve $\alpha C_\Omega$ clicks, a uniform strategy must randomize between bids $h(u)$ and $h(v)$ where $u \leq \alpha C_\Omega$ and $v \geq \alpha C_\Omega$. Call the set of such strategies $S$. Given a $(u, v) \in S$, the necessary probabilities in order to achieve $\alpha C_\Omega$ clicks are easily determined, and we denote them by $p_1(u, v)$ and $p_2(u, v)$ respectively. Note further that the advertiser is trying to figure out which of these strategies to use, and ultimately wants to compute a distribution over uniform strategies. In the LP, she is actually going to compute a distribution over pairs of strategies in $S$, which we will then interpret as a distribution over strategies.

Using this set of uniform strategies as constraints, we can characterize a set of worst-case click-price curves by the constraints

$$\int_0^{C_\Omega} h(r)dr \leq U$$

$$\forall (u,v) \in S \quad p_1(u,v)uh(u) + p_2(u,v)vh(v) \geq U$$

A curve $h$ that satisfies these constraints has the property that all uniform strategies that obtain $\alpha C_\Omega$ clicks spend more than $U$. Discretizing this set of inequalities, and pushing the first constraint into the objective function, we get the following LP over variables $h_r$ representing the curve:

$$\min \sum_{r \in \{0, \epsilon, 2\epsilon, \ldots, C_\Omega\}} \epsilon \cdot h_r \text{ s.t.}$$

$$\forall (u,v) \in S, \quad p_1(u,v)uh_u + p_2(u,v)vh_v \geq U$$

In this LP, $S$ is defined in the discrete domain as $S = \{(u,v) \in \{0, \epsilon, 2\epsilon, \ldots, C_\Omega\}^2 : 0 \leq u \leq \alpha C_\Omega \leq v \leq C_\Omega\}$.

Solving this LP for a particular $\alpha$, if we get an objective less than $U$, we know (up to some discretization) that an instance of BUDGET OPTIMIZATION exists that cannot be approximated better than $\alpha$. (The instance is constructed as in the proof of Lemma 5.) A binary search yields the lowest such $\alpha$ where the objective is exactly $U$.

To obtain a strategy for the advertiser, we look at the dual, constraining the objective to be equal to $U$ in order to get the polytope of optimum solutions:

$$\sum_{(u,v) \in S} w_{u,v} = 1$$

$$\forall (u,v) \in S, \quad \sum_{v':(u,v') \in S} p_1(u,v')uw_{u,v'} \leq \epsilon \text{ and}$$

$$\sum_{u':(u',v) \in S} p_2(u',v)vw_{u',v} \leq \epsilon.$$

It is straightforward to show that the second set of constraints is equivalent to the following:

$$\forall h \in \mathbb{R}^{C_\Omega/\epsilon} : \sum_r \epsilon h_r = U,$$

$$\sum_{(u,v) \in S} w_{u,v}(p_1(u,v)uh_u + p_2(u,v)vh_v) \leq U.$$

Here the variables can be interpreted as weights on strategies in $S$. A point in this polytope represents a convex combination over strategies in $S$, with the property that for *any* click-price curve $h$, the cost of the mixed strategy is at most $U$. Since all strategies in $S$ get at least $\alpha C_\Omega$ clicks, we have a strategy that achieves an $\alpha$-approximation. Interestingly, the equivalence between this polytope and the LP dual above shows that there is a mixture over values $r \in [0, C]$ that achieves an $\alpha$-approximation for *any curve* $h$.

After a search for the appropriate $\alpha$ (which turned out to be $1 - \frac{1}{e}$), we solved these two LPs and came up with the plots in Figure 4 (see appendix), which reveal not only the right approximation ratio, but also a picture of the worst-case distribution and the approximation-achieving strategy.[10] From the pictures, we were able to quickly "guess" the optimal strategy and worst case example.

**4.3.2 Proof of Theorem 3.** We can now complete the proof of Theorem 3.

*Proof of Theorem 3:* By Lemma 4, we know that for each $r \leq U_\Omega$, there is a strategy that can obtain traffic $r$ at cost $r \cdot h(r)$. By mixing strategies for multiple values of $r$, we construct a uniform strategy that is guaranteed to achieve at least $1 - e^{-1} = 0.63$ fraction of $\Omega$'s traffic and remain within budget. Note that the "final" resulting bid distribution will have some weight on the zero bid, since the single-bid strategies from Lemma 4 put some weight on bidding zero.

Consider the following probability density function over such strategies (also depicted in Figure 4):

$$g(r) = \begin{cases} 0 & \text{for } r < C_\Omega/e \\ 1/r & \text{for } r \geq C_\Omega/e \end{cases}$$

Note that $\int_0^{C_\Omega} g(r)\,dr = \int_{C_\Omega/e}^{C_\Omega} \frac{1}{r}dr = 1$, i.e. $g$ is a probability density function. The traffic achieved by our strategy is equal to

$$\text{traffic} = \int_0^{C_\Omega} g(r) \cdot r\, dr = \int_{C_\Omega/e}^{C_\Omega} \frac{1}{r} \cdot r\, dr = \left(1 - \frac{1}{e}\right) C_\Omega.$$

The expected total spend of this strategy is at most

$$\text{spend} = \int_0^{C_\Omega} g(r) \cdot rh(r)\, dr$$

$$= \int_{C_\Omega/e}^{C_\Omega} h(r)\, dr \leq \int_0^{C_\Omega} h(r)\, dr = U_\Omega.$$

---
[10]The parameters $U$ and $C_\Omega$ can be set arbitrarily using scaling arguments.

Thus we have shown that there exists a uniform bidding strategy that is $(1 - \frac{1}{e})$-optimal.

We now show that no uniform strategy can do better. We will prove that for all $\epsilon > 0$ there exists an instance for which all uniform strategies are at most $(1 - \frac{1}{e} + \epsilon)$-optimal.

First we define the following click-price curve over the domain $[0, 1]$:

$$h(r) = \begin{cases} 0 & \text{for } r < e^{-1} \\ \frac{1}{e-2}\left(e - \frac{1}{r}\right) & \text{for } r \geq e^{-1} \end{cases}$$

Note that $h$ is non-decreasing and non-negative. Since the curve is over the domain $[0, 1]$ it corresponds to an instance where $C_\Omega = 1$. Note also that $\int_0^1 h(r)\,dr = \frac{1}{e-2}\int_{1/e}^1 e - \frac{1}{r}\,dr = 1$. Thus, this curve corresponds to an instance where $U_\Omega = 1$. Using Lemma 5, we construct such an instance where the best uniform strategies are convex combinations of strategies that bid $h(u)$ and achieve $u$ clicks and $u \cdot h(u)$ cost.

Suppose for the sake of contradiction that there exists a uniform bidding strategy that achieves $\alpha > 1 - e^{-1}$ traffic on this instance. By Lemma 1 there is always a two-bid optimal uniform bidding strategy and so we may assume that the strategy achieving $\alpha$ clicks randomizes over two bids. To achieve $\alpha$ clicks, the two bids must be on values $h(u)$ and $h(v)$ with probabilities $p_u$ and $p_v$ such that $p_u + p_v = 1$, $0 \leq u \leq \alpha \leq v$ and $p_u u + p_v v = \alpha$.

To calculate the spend of this strategy consider two cases: if $u = 0$ then we are bidding $h(v)$ with probability $p_v = \alpha/v$. The spend in this case is:

$$\text{spend} = p_v \cdot vh(v) = \alpha h(v) = \frac{\alpha e - \alpha/v}{e - 2}$$

Using $v \geq \alpha$ and then $\alpha > 1 - \frac{1}{e}$ we get

$$\text{spend} \geq \frac{\alpha e - 1}{e - 2} > \frac{(1 - 1/e)e - 1}{e - 2} = 1,$$

contradicting the assumption.

We turn to the case $u > 0$. Here we have $p_u = \frac{v - \alpha}{v - u}$ and $p_v = \frac{\alpha - u}{v - u}$. Note that for $r \in (0, 1]$ we have $h(r) \geq \frac{1}{e-2}(e - \frac{1}{r})$. Thus

$$\begin{aligned} \text{spend} &\geq p_u \cdot uh(u) + p_v \cdot vh(v) \\ &= \frac{(v - \alpha)(ue - 1) + (\alpha - u)(ve - 1)}{(v - u)(e - 2)} \\ &= \frac{\alpha e - 1}{e - 2} \\ &> 1 \end{aligned}$$

The final inequality follows from $\alpha > 1 - \frac{1}{e}$. Thus in both cases the spend of our strategy is over the budget of 1. ∎

## 5 Extensions and Concluding Remarks

The BUDGET OPTIMIZATION problem can be solved exactly for several special cases, first when the graph is a matching, then when we know an ordering on the keyword bids and finally on a more general scenario in which the keywords define a laminar family of sets over the queries. These results appear in Appendix B.

In its generality, the BUDGET OPTIMIZATION problem is hard. It is trivially weakly NP-hard because it encodes the Knapsack problem. We show further that the BUDGET OPTIMIZATION problem is strongly NP-hard, and a weighted generalization of the BUDGET OPTIMIZATION problem is hard to approximate to within $(1 - 1/e)$. These results appear in Appendix C.

We have done some preliminary experiments using real auction data logs at Google. The uniform bidding strategies seem to be significantly better than the theoretical guarantees of 0.5 and 0.63. We plan to do more detailed experimental study and also understand the properties of real inputs.

Our algorithmic result presents an intriguing heuristic in practice. Say an advertiser chooses some bid value $b$ and bids that on all keywords. At the end of the day, if the budget is underspent, they adjust $b$ to be higher; if budget is overspent, they adjust $b$ to be lower; else, they maintain $b$. If the scenario does not change from day to day, this simple strategy will have the same theoretical properties as our one-bid strategy, and in practice, likely to be much better. Our belief is that this heuristic may eventually become one of choice for advertisers since it does not involve detailed consideration of landscapes and interactions. We intend to evaluate this scheme experimentally.

## Acknowledgments

We thank Rohit Rao, Zoya Svitkina and Adam Wildavsky for helpful discussions.

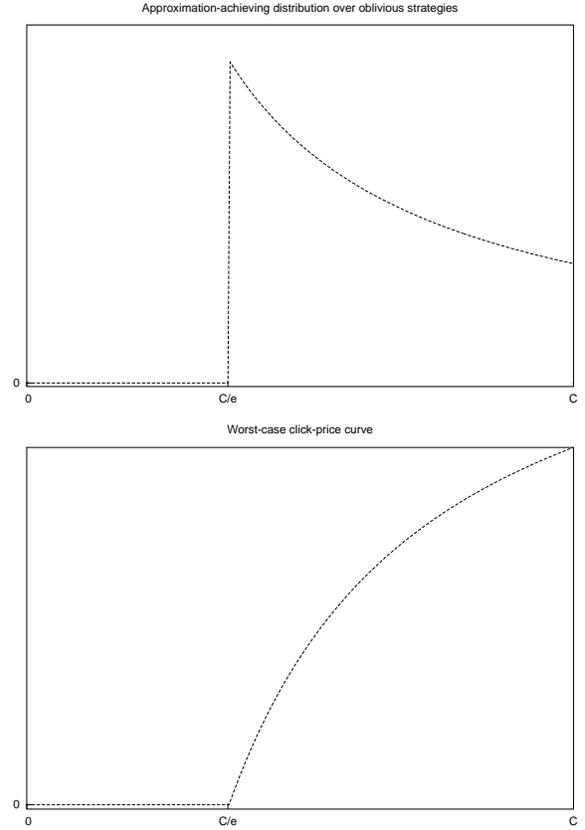

Figure 4: The worst-case click-price curve and $(1 - 1/e)$-approximate uniform bidding strategy, as found by linear programming.

## A  Proofs

*Proof of Lemma 5:* Construct an instance as follows. Let $\epsilon > 0$ be a small number that we will later define in terms of $\epsilon'$. Define $r_0 = 0, r_1, r_2, \ldots, r_m = C$ such that $r_{i-1} < r_i \leq r_{i-1} + \epsilon$, $f(r_{i-1}) \leq f(r_i) \leq f(r_{i-1}) + \epsilon$, and $m \leq (C + f(C))/\epsilon$. (This is possible by choosing $r_i$'s spaced by $\min(\epsilon, f(r_i) - f(r_{i-1}))$) Now make a query $q_i$ for all $i \in [m]$ with bidders bidding $h(r_i), h(r_{i+1}), \ldots, h(r_m)$, and ctr's $\alpha[1] = \alpha[2] = \cdots = \alpha[m-i+1] = r_i - r_{i-1}$. The graph is a matching with one keyword per query, and so we can imagine the optimal solution as bidding on queries. The optimal solution will always bid exactly $h(r_i)$ on query $q_i$, and if it did so on all queries, it would spend $U' := \sum_{i=1}^m (r_i - r_{i-1}) h(r_i)$. Define $\epsilon$ small enough so that $U' = U + \epsilon'$, which is always possible since

$$\begin{aligned} U' &\leq \int_0^C f(r)dr + \sum_{i=1}^m (r_i - r_{i-1})(h(r_i) - h(r_{i-1})) \\ &\leq U + \epsilon^2 m \leq U + \epsilon(C + f(C)). \end{aligned}$$

Note that the only possible bids (i.e., all others have the same results as one of these) are $f(r_0), \ldots, f(r_m)$, and bidding uniformly with $f(r_i)$ results in $\sum_{j=1}^i r_j - r_{i-1} = r_i$ clicks at cost $h(r_i)r_i$. ∎

*Proof of Theorem 6:* Here we complete the proof of Theorem 6. For the second part of the theorem, to construct a tight example, we use two queries $Q = \{x, y\}$, two keywords $K = \{u, v\}$, and edges $E = \{(u, x), (v, y)\}$.

Fix some $\alpha$ where $0 < \alpha \leq 1$, and fix some very small $\epsilon > 0$. Query $x$ has two positions, with bids of $b_1^x = 1/\alpha$ and $b_2^x = \epsilon$, and with identical click-through rates $\alpha^x[1] = \alpha^x[2] = \alpha$. Query $y$ has one position, with a bid of $b_1^y = 1/\alpha$ and a click-through rate of $\alpha^y[1] = \alpha$. The budget is $U = 1 + \epsilon\alpha$. The optimal solution is to bid $\epsilon$ on $u$ (and therefore $x$) and bid $1/\alpha$ on $v$ (and therefore $y$), both with probability 1. This achieves a total of $2\alpha$ clicks and spends the budget exactly. The only useful bids are 0, $\epsilon$ and $1/\alpha$, since for both queries all other bids are identical in terms of cost and clicks to one of those three. Any single-bid solution that uses $\epsilon$ as its non-zero bid gets at most $\alpha$ clicks. Bidding $1/\alpha$ on both keywords results in $2\alpha$ clicks and total cost 2. Thus, since the budget is $U = 1 + \epsilon\alpha < 2$, a single-bid solution using $1/\alpha$ can put weight at most $(1 + \epsilon\alpha)/2$ on the $1/\alpha$ bid. This results in at most $\alpha(1 + \epsilon\alpha)$ clicks. This can be made arbitrarily close to $\alpha$ by lowering $\epsilon$. ∎

## B  Beyond Uniform Strategies: optimal strategies for some graphs

In this section we solve the BUDGET OPTIMIZATION problem exactly for several special cases, first when the graph is a matching, then when we know an ordering on the keyword bids and finally on a more general scenario in which the keywords define a laminar family of sets over the queries. We will also show how the matching defines an upper bound, useful in later sections.

**B.1  A matching.** We consider a graph $G$ which is a matching, that is, each node has degree at most 1. In this case, the advertiser can bid on each query independently (via the unique matched keyword), and the interaction among keywords is not an issue. The resulting optimization problem is a small generalization of the *fractional knapsack* problem in which each item has a *size* and *value* and we wish to select a set of items such that the total size is at most the knapsack size while maximizing the total value. We now sketch an algorithm, due to Labio et al. [8].

The first step of the algorithm is to take the convex hull of each query landscape, as in Figure 2, and remove any landscape points not on the convex hull. (See the discussion around Lemma 1 for justification.) Then, for each query we have a piecewise linear concave curve. Examining each "piece" of these curves more carefully, we see that a particular piece represents the incremental number of clicks and cost incurred by moving one's bid on a keyword (and therefore its matching query) from one particular bid to another. More precisely, for the piece connecting two consecutive bids $b'$ and $b''$ on the convex hull, we create an entry with cost $\text{cost}(b'') - \text{cost}(b')$ and clicks $\text{clicks}(b'') - \text{clicks}(b')$. These values represent the incremental cost and clicks associated with moving from bid $b'$ to bid $b''$. For example, in Figure 2, the piece between the second and third point represents getting .25 clicks at cost \$0.80 by moving one's bid from 0.50 to \$2.00.

We then regard these "pieces" as items in an instance of fractional knapsack with value equal to the incremental number of clicks and size equal to the incremental cost. We emulate the greedy algorithm for knapsack, sorting by cost-per-click, and choosing greedily until the budget is exhausted. It is well-known that in the optimal solution, at most one item is chosen fractionally. Choosing an item fractionally in this situation corresponds to bidding randomly between the two bid values that define the line segment.

In this reduction to knapsack we have ignored the fact that some of the pieces come from the same query and cannot be treated independently. However, since each curve is concave, the pieces that come from a particular query curve are in increasing order of cost-per-click; thus from each query we have chosen for our "knapsack" a set of pieces that form a prefix of the curve. Hence for each query there is a bid (or in one case a convex combination of two bids) that exactly achieves cost and clicks equal to the sum of the costs and clicks implied by the pieces.

**Lemma 7** *If the graph $G$ is a matching, then the greedy algorithm described above solves the BUDGET OPTIMIZATION problem exactly in polynomial time.*

**B.2 A star graph.** We next consider the case when the graph G is a star, with only one keyword node that is connected to every query node. In this case we only make one bid, and that bid is applied to all queries. This is equivalent to finding the best uniform strategy in a general graph, and so can be done in polynomial time by Lemma 2.

**B.3 A disjoint union of star graphs.** Combining the results of the previous two sections, we have an algorithm for the BUDGET OPTIMIZATION problem for the case when $G$ is a disjoint union of star graphs. For each star, if the center of the star is a keyword, we form the aggregate landscape for that star, as in the previous section. If the center of a star is a query, we just pick one of the keywords arbitrarily and delete the edges to the remaining keywords. After this transformation, we have a matching, and can apply Lemma 7

**Lemma 8** *If the graph G is a disjoint union of star graphs, then it can be transformed into an matching problem with aggregate landscapes and solved exactly in polynomial time.*

**B.4 Laminar structure in graphs.** We now consider several cases where the graph has a laminar structure. As we shall see, this structure will allow us to impose a (partial) ordering on the possible bid values, and thereby give pseudopolynomial algorithms via dynamic programming.

Before doing this, we show that to solve the BUDGET OPTIMIZATION problem (for general graphs) optimally in pseudopolynomial time, it suffices to provide an algorithm that solves the deterministic case.

**Lemma 9** *Let I be an input to the BUDGET OPTIMIZATION problem and suppose that we find the optimal deterministic solution for every possible budget $U' \leq U$. Then we can find the optimal solution in time $O(U \log U)$.*

*Proof:* Let $\text{traffic}(U')$ be the optimal number of clicks given a budget of $U'$. We can think of the pair $(U', \text{traffic}(U'))$ as a point on a landscape, with cost $U'$ and clicks $\text{traffic}(U')$. Properties analogous to those in Observation 1 now apply to the set $V = \{(U', \text{traffic}(U')) : U' \leq U\}$, i.e. cost and clicks are non-decreasing, as is cost per click. We can thus apply Lemma 1 to show that the optimal distribution consists of 2 points on the convex hull of this landscape. Finding the convex hull takes time $O(U \log U)$[10]. ∎

Given a collection of $n$ sets $S_1, \ldots, S_2$, we say that $S$ is *nested* if $S_1 \subseteq S_2 \subseteq \cdots \subseteq S_n$. We say that $S$ is *laminar* if, for any two sets $S_i$ and $S_j$, if $S_i \cap S_j \neq \emptyset$ then either $S_i \subseteq S_j$ or $S_j \subseteq S_i$.

Given a keyword interaction graph $G$, we associate a set of neighboring queries $Q_k = \{q : (k, q) \in E\}$ with each keyword $k$. If this collection of sets is nested, we say that the graph has the *nested property* and if the collection of sets if laminar, we say that the graph has the *laminar property*.

The following lemma will be useful for giving a structure to the optimal solution, and will allow a dynamic programming approach to go through.

We call a solution *deterministic* if it consists of one bid vector, rather than a general distribution over bid vectors.

**Lemma 10** *For keywords $i, j \in K$, if $Q_i \subseteq Q_j$ then there exists an optimal deterministic solution to the BUDGET OPTIMIZATION problem with $\mathbf{a}_i \geq \mathbf{a}_j$.*

*Proof:* Assume that some optimal solution had $\mathbf{a}_i < \mathbf{a}_j$. Because we assign queries to the highest incident keyword, all queries in $Q_i$ are assigned either to keyword $j$ at bid level $\mathbf{a}_j$ or to some other keyword $j'$ at bid level greater than $\mathbf{a}_j$. Therefore, we can increase $\mathbf{a}_i$ to be equal to $\mathbf{a}_j$ without changing the effective bid of any keyword. ∎

**B.4.1 Nested Sets.** We first consider the case when the sets $Q_i$ of queries are nested, with $Q_1 \subseteq Q_2 \subseteq \cdots \subseteq Q_n$. Note that by Lemma 10, we can assume that the bids are ordered with $\mathbf{a}_1 \geq \mathbf{a}_2 \geq \cdots \geq \mathbf{a}_n$, and further, the set of queries that will be assigned to keyword $i$ is exactly the set $Q_i \setminus Q_{i-1}$. We can, therefore, given a keyword $i$ and a bid $\mathbf{a}_i$, compute an incremental spend and traffic associated with bidding $\mathbf{a}_i$ on keyword $i$, that is

$$\hat{t}(i, \mathbf{a}_i) = \sum_{q \in Q_i \setminus Q_{i-1}} \text{clicks}_q(\mathbf{a}_i), \quad \text{and}$$

$$\hat{s}(i, \mathbf{a}_i) = \sum_{q \in Q_i \setminus Q_{i-1}} \text{cost}_q(\mathbf{a}_i).$$

We define $F[i, b, U]$ to be the maximum number of clicks achievable by bidding at least $b$ on each of keywords $1, \ldots, i$ (and exactly $b$ on keyword $i$) while spending at most $U$. We now give a recurrence for $F[i, b, U]$ by considering all possible minimum bids and budgets for the set of keywords $1, \ldots, i-1$. For each such choice, we consider bidding $b$ on keyword $i$ yielding some number of incremental clicks and incremental spend. If the incremental spend is at most $U - U'$, then we have a new candidate solution equal to $F[i-1, b', U'] + \hat{t}(i, b)$. If the incremental spend would exceed the budget then we return a 0. (We also return a zero for the base case of $i = 0$. That is, we have $F[i, b, U] =$

$$\max_{b' \geq b,\ U' \leq U} \{F[i-1, b', U'] + \hat{t}(i, b)\} \quad (5)$$

if $(\hat{s}(i, b) \leq U - U'$ and $i > 0)$, and $F[i, b, U] = 0$ otherwise.

Assuming that budgets take on integer values of at most $U$ and bids take on integer values of at most $B$, we have the following

**Lemma 11** *If the graph $G$ has the nesting property, then the dynamic programming recurrence in (5) finds an optimal deterministic solution to the* BUDGET OPTIMIZATION *problem exactly in $O(B^2 U^2 n)$ time.*

*Proof:* The table is of size $nBU$ and each update requires a max over $BU$ values, each of which can be computed in constant time. ∎

Using standard rounding and grouping ideas, we can compute an approximately optimal solution in polynomial time. We omit the details, which are essentially subsumed by the laminar set case below.

**B.5 Laminar Sets.** We can extend the ideas used for graphs with the nesting property to graphs with the laminar property. We still use Lemma 10 to restrict the bids on the sets. We can view the laminar order as a tree with keyword $j$ as a parent of keyword $i$ if $Q_j$ is the minimal set containing $Q_i$. In this case we say that $j$ is a child of $i$. Given a keyword $j$ with $c$ children $i_1, \ldots, i_c$, we now need to enumerate over all ways to allocate the budget among the children and also over all possible minimum bids for the children. A serious complication is that a node may have many children and thus a term of $U^c$ would not even be pseudopolynomial. We can solve this problem by showing that given any laminar ordering, there is an equivalent one in which each keyword has at most 2 children.

**Lemma 12** , *Let $G$ be a graph with the laminar property. There exists another graph $G'$ with the same optimal solution to the* BUDGET OPTIMIZATION *problem, where each node has at most two children in the laminar ordering. Furthermore, $G'$ has at most twice as many nodes as $G$.*

*Proof:* We will accomplish this transformation by adding "dummy" keyword nodes. Suppose we have two nodes $i$ and $j$ that are siblings in the tree, with parent $p$. Now add a new keyword $\ell$, with $Q_\ell = Q_i \cup Q_j$. Trivially, we must also have $Q_\ell \subseteq Q_p$, and $Q_\ell \backslash (Q_i \cup Q_j) = \emptyset$. Now consider the optimal solution to our transformed problem. If $\mathbf{a}_\ell \geq \min\{\mathbf{a}_i, \mathbf{a}_j\}$, we can increase $\mathbf{a}_i$ and/or $\mathbf{a}_j$ to be equal to $\mathbf{a}_\ell$ without changing the effective bid on any of our queries. We thus have established that $\mathbf{a}_\ell \leq \min\{\mathbf{a}_i, \mathbf{a}_j\}$, and hence keyword $\ell$ does not get assigned any queries and can be removed. ∎

Given a graph with at most two children per node, we can extend (5). We redefine $F[i, b, U]$ to be the maximum number of clicks achievable by bidding at least $b$ on each of keywords $j$ s.t. $Q_j \subseteq Q_i$ (and exactly $b$ on keyword $i$) while spending at most $U$. For this definition, we use $Z(b, U)$ to denote set of allowable bids and budgets over children:

$$\begin{aligned} Z(b, U) &= \{b', b'', U', U'' : b' \geq b, U' \leq U, \\ &\quad b'' \geq b, U'' \leq U, U' + U'' \leq U\} \end{aligned}$$

Now we define $F[i, b, U]$ as

$$\max_{\substack{b', b'', U', U'' \\ \in Z(b, U)}} \left\{ F[j', b', U'] + F[j'', b'', U''] + \hat{t}(Q_i, b) \right\} \quad (6)$$

if $(\hat{s}(Q_i, b) \leq U - U' - U''$ and $i > 0)$, and $F[i, b, U] = 0$ otherwise.

By a proof similar to that of Lemma 11, we have

**Lemma 13** *If the graph $G$ has the laminar property, then, after applying Lemma 12, the dynamic programming recurrence in (6) finds an optimal deterministic solution to the* BUDGET OPTIMIZATION *problem exactly in $O(B^3 U^3 n)$ time.*

In addition, we can apply Lemma 9 to compute the optimal (randomized) solution. Observe that in the dynamic program, we have already solved the instance for every budget $U' \leq U$, so we can find the randomized solution with no additional asymptotic overhead.

**Lemma 14** *If the graph $G$ has the laminar property, then, by applying Lemma 12, the dynamic programming recurrence in (6), and Lemma 9, we can find an optimal solution to the* BUDGET OPTIMIZATION *problem in $O(B^3 U^3 n)$ time.*

The bounds in this section make pessimistic assumptions about having to try *every* budget and *every* level. For many problems, you only need to choose from a discrete set of bid levels (e.g., multiples of one cent). Doing so yields the obvious improvement in the bounds.

## C Hardness

In this section we prove that the BUDGET OPTIMIZATION is strongly NP-hard, and that a generalization of it is in fact $(1 - \frac{1}{e})$-hard.

### C.1 Strong NP-hardness.

**Theorem 15** *The* BUDGET OPTIMIZATION *problem is strongly NP-hard.*

*Proof:* (sketch) By reduction from Vertex Cover. Let $[H = (V_H, E_H), k^*]$ be an instance of vertex cover, where $H$ is the graph and $k^*$ is the size of the vertex cover. We make an interaction graph $G = (K \cup Q, E)$ where $|K| = |V_H|$, and we have $Q = R \cup S$ where $|R| = |E_H|$ and $|S| = |V_H|$. We make an edge $(k, q) \in E$ between $k \in K$ and $q \in R$ if (in $H$) the vertex associated with $k$ is incident to the edge associated with $q$. For the query set $S$ we add edges to make a matching between $S$ and the keywords $K$. To complete our instance of BUDGET OPTIMIZATION it remains to define landscapes for each query. For the queries $s \in S$, we have two positions with ctrs $\alpha^s[1] = \alpha^s[2] = 1.0$, and bids $b_1^s = \$\epsilon$ and $b_2^s = \$1$. For queries $r \in R$ we have one position with ctr $\alpha^r[1] = 1.0$, and there is one bid $b_1^r = \$1$. We set the budget equal to $(|V_H| - k^*)\epsilon + |E_H| + k^*$.

We claim that there is a vertex cover of size $k^*$ if and only if there is a solution to this instance that achieves $|E_H| + |V_H|$ clicks.

For the forward direction, we bid \$1 on every keyword associated with a vertex in the cover, and \$$\epsilon$ on every other vertex. We win the click on every query in $S$, and since this is a vertex cover, we win the click on every query in $R$, for a total of $|E_H| + |V_H|$ clicks. Our cost is \$1 for each $r \in R$, \$1 for each $s \in S$ incident to a keyword associated with a vertex in the cover, and \$$\epsilon$ for all other $s \in S$, for a total of exactly the budget.

For the backward direction, we note that the total number of clicks available is $|E_H| + |V_H|$ and thus all of them must be won. This implies that the keywords on which we bid \$1 must form a cover of the graph $H$. The total cost incurred by these keywords is $|E_H| + k'$, where $k'$ is the size of the cover. Making $\epsilon$ small enough (say $|E_H|^{-2}$), we see that we must get $k' \leq k^*$ in order to remain within budget. ∎

### C.2 Hardness of approximation (with click values).

Suppose we introduce weights on the queries that indicate the relative value of a click from the various search users. Formally, we have weights $w_q$ for all $q \in Q$ and our goal is maximize the total *weighted* traffic given a budget. Call this the WEIGHTED KEYWORD BIDDING problem.

With this additional generalization we can show hardness of approximation via a simple reduction from the MAXIMUM COVERAGE problem, which is know to be $(1 - 1/e)$-hard [5]. In this problem you are given a collection $S_1, \ldots, S_m$ of sets over $[n]$ and a number $k^*$, and you are asked to find a set $I \subset [m]$ of size $k^*$ that maximizes $| \cup_{i \in I} S_i|$, the number of covered items.

**Theorem 16** *The* WEIGHTED KEYWORD BIDDING *problem is hard to approximate to within* $(1 - 1/e)$.

*Proof:* (sketch) By approximation-preserving reduction from MAXIMUM COVERAGE. Given an instance $[(S_1, \ldots, S_m), k^*]$ of MAXIMUM COVERAGE, we create an instance of WEIGHTED KEYWORD BIDDING with $|K| = m$, and $Q = R \cup S$, where $|R| = n$ and $|S| = m$. We create s correspondence between keywords $k \in K$ and sets $S_i$, and one between queries $q \in R$ and found elements $e \in [n]$. Now, if $S_i \ni e$, then we connect the corresponding keyword and query. We also add a matching between keywords in $K$ and queries in $S$. This defines the keyword graph. For the queries $q \in R$, there is one position with $\alpha[1] = 1.0$, and one bidder bidding \$1. Additionally, we set $w_q = 1$. For the queries $q \in S$, there is one position with $\alpha[1] = n^2$, and one bidder bidding \$1.[11] For these queries we set the weight $w_q = 0$. The budget is set to $k^* n^2 + n$.

It is simple to show that an algorithm that can approximate this instance of WEIGHTED KEYWORD BIDDING to within $\gamma$ also approximates the MAXIMUM COVERAGE instance to within $\gamma$. ∎

---

[11]To simulate a ctr of $n^2$ we could make $n^2$ identical copies of the query each with ctr 1.0

15